# Simultaneous detection of the size and velocity of the largest ejecta particles with velocities exceeding 1 km $s^{-1}$

Short title: Size/velocity detection of high-speed ejecta

Akiko M. Nakamura[1,2], Keita Nomura[2], Sunao Hasegawa[3]

Abstract

Impact ejecta with velocities exceeding the escape velocity of planetary bodies become meteorites and dust particles in interplanetary space. We present a new method that allows simultaneous measurement of the size and velocity of the largest high-velocity ejecta. High-speed camera images revealed the time required for the ejecta to reach the secondary target, and ejecta size was determined after the experiment by analyzing the craters formed upon their impact on the secondary target. We defined the size-velocity relationships of sub-millimeter ejecta with velocities exceeding 1 km $s^{-1}$, focusing on the largest detectable ejecta in our experiments. The results show that millimeter-sized meteoroids impacting the rocky surfaces of planetary bodies at 7 km $s^{-1}$ eject particles up to a few tens of micrometers in size toward interplanetary space at velocities exceeding the escape velocity of the body, even when it is greater than 1 km $s^{-1}$.

[1] Corresponding author amnakamu@kobe-u.ac.jp

[2] Department of Planetology, Kobe University, 1-1 Rokkodai-cho, Nada-ku, Kobe-city, Hyogo 657-8501, Japan

[3] Institute of Space and Astronautical Science, Japan Aerospace Exploration Agency, 3-1-1 Yoshinodai, Chuo-ku, Sagamihara-city, Kanagawa 252-5210, Japan

# 1. Introduction

Hypervelocity impacts on planetary bodies are followed by the ejection of materials from their surfaces. The size–velocity distribution of ejecta is a fundamental outcome of such impacts and provides an essential constraint on the collisional evolution of planetary systems and the impact processes that shape planetary and small bodies.

Ejecta that do not reach the escape velocity of a body fall back to the surface, excavating it to form secondary craters, reaccumulating to form blocks and regolith deposits (e.g., Hirata & Nakamura 2006; Bart & Melosh 2010), or flowing into closed topographic depressions (e.g., Cao et al., 2024). The secondary craters formed by the re-impact of ejecta on the surface are an important factor to consider in crater chronology (Bierhaus et al. 2018; Xiao 2018). The reaccumulation process produces characteristic surface features, such as lunar rock deposits that accumulate abundantly on slopes with specific orientations depending on the direction of ejecta arrival (Bandfield et al. 2017), and may have contributed to the formation of the equatorial ridge on asteroid Ryugu (Ikeya & Hirata 2021).

In contrast, ejecta that exceed the escape velocity of the body reach interplanetary space (Ishiguro et al. 2011; Kim & Jewitt 2023) where they may eventually fall to Earth as lunar or Martian meteorites (e.g., Melosh 1984; Warren 1994; Head et al. 2002; Artemieva & Ivanov 2004; Hyodo & Genda 2018) or contribute to circumplanetary and interplanetary dust populations (Szalay et al. 2018; Kruger et al. 2024).

High-resolution imaging of blocks and secondary craters on the Moon, Mars, Mercury, and the Jovian satellites, Europa and Ganymede, has provided information on the size–velocity distribution of ejecta blocks with velocities from several tens to several hundreds of meters per second (Vickery 1986; 1987; Hirase et al. 2004; Bart & Melosh 2010; Singer et al. 2013; 2020; Xu et al. 2023; 2024), even up to 1.8 km $s^{-1}$ on the Moon (Guo et al. 2018). In these studies, ejecta velocities were estimated from the distances to the secondary craters, assuming both the launch angle and the launch point within the primary crater. The sizes of ejecta forming secondary craters were inferred from the diameters of secondary craters and the empirical crater-scaling relationships applicable to the gravity- or strength-dominated regimes (Holsapple 1993). Such analyses of ejecta blocks and secondary craters yield ejecta size–velocity relationships for impacts much larger than those achievable in laboratory experiments. However, these estimates are limited to ejecta with velocities below the escape velocity of the object of which the images are taken.

In high-velocity impact experiments, ejecta cloud with velocities comparable to the impact velocity of a projectile was captured by high-speed camera imaging at rates of $5\times10^6$ frames per second (fps) and compared with numerical simulation results (Okamoto et al. 2020). From the analysis of high-speed camera photographs taken at rates of up to $10^6$ fps, the relationship between ejecta velocity and ejection angle, as well as the integrated mass versus velocity were derived (Gault 1963). However, these studies did not investigate the relationship between the size of individual ejecta and their velocities.

For ejecta with velocities lower than 100 m $s^{-1}$, both ejecta size and velocity were directly obtained from high-speed camera images (Nakamura & Fujiwara 1991; Nakamura 1993; Giblin et al. 2004; Onose & Fujiwara 2004). Ejecta with velocities > 1 km $s^{-1}$ are typically sub-millimeter-sized or smaller. Such a small size, combined with rapid motion, makes image tracking and sizing particularly challenging. For example, a particle traveling at a velocity of 5 km $s^{-1}$ moves 5 mm during an imaging interval of 1 µs. If the camera's field of view is 20 mm with 400 pixels, the spatial resolution is 50 µm $pixel^{-1}$. At this resolution, particles smaller than ~100 µm cannot be resolved. Additionally, without a powerful light source and a very short exposure time, the image appears blurred in the direction of the ejecta's movement, further complicating the determination of particle size. Thus, the lower velocity limits of small ejecta (> 100 m s-1) have been investigated using secondary targets (thin aluminum and plastic sheets and films with thicknesses ranging from 0.8 to 30 µm) (Nakamura et al. 1994). In that study, the smallest velocity capable of penetrating the film or foil was determined to be the lower limit of ejecta velocity. On the other hand, the ejecta size was assumed to be equal to the hole made in the film or foil. This assumption appears to be valid for penetration velocities on the order of ~km $s^{-1}$ and for ejecta sizes that are several tens of times greater than the thickness of the foil or film, based on previous studies on penetration. However, for ejecta sizes close to the thickness of the foil or film, a hole larger than the ejecta is formed, which has been shown to depend on velocity (e.g., Hoerz 2012). Thus, that method provides only the lower limit of the ejecta speed and incomplete information regarding the size-velocity relationship.

Here, we present a new method that simultaneously determines the size and velocity of fast impact ejecta. The arrival of ejecta particles at a secondary target is associated with changes in grayscale of pixels in images from high-speed cameras; the time of such changes yields the ejecta particle velocity. Then the size of each particle is determined by measuring the size of the crater in the microscope image of the secondary target, where the time of impact is detected by the high-speed camera. Using the velocity information obtained earlier, this size is substituted into the crater scaling law for the secondary target to determine the size of each ejecta particle that produced the crater. We describe the procedure in detail and compare the results to those of previous laboratory studies.

## 2. Impact Experiments

### *2.1. Impact Experiments to Observe Ejecta*

The two-stage light-gas gun of the Institute of Space and Astronautical Science, Japan Aerospace Exploration Agency (ISAS, JAXA) was employed. Table 1 summarizes the experimental settings. Figure 1 shows the experimental setup.

Spherical projectiles of aluminum alloy (5052) and alumina 3.2 mm in diameter were accelerated using a split-type nylon sabot (Kawai et al. 2010) to impact a target surface perpendicularly (i.e., $\alpha$=90° in Fig. 1a) at 7 km s$^{-1}$, which is the maximum velocity typically achievable with this gun. The projectile speed was determined from the flight time between two laser beams and the interval of the beams.

Table 1. Experimental conditions

| Shot# | Projectile | | | Secondary target | | Cam-1 setting | | Cam-2 setting | | |
|---|---|---|---|---|---|---|---|---|---|---|
| | Material | Mass | Impact velocity | $r_0$[1] | $\theta_0$[2] | $T_{int}$[3] | $T_{exp}$[4] | $T_{int}$[3] | $T_{exp}$[4] | Resolution[5] |
| | … | (g) | (km s$^{-1}$) | (mm) | (°) | (μs) | (μs) | (μs) | (μs) | (mm pixel$^{-1}$) |
| 4982 | Al | 0.0450 | 6.86 | 370 | 58 | 2 | 0.5 | 33.3 | 2 | 0.44 |
| 5115 | Alumina | 0.0651 | 6.80 | 440 | 57 | 2 | 0.25 | 10 | 2 | 0.14 |
| | | | | | | | | 10 | 8 | 0.41 |
| 5231 | Al | 0.0451 | 6.96 | 400 | 60 | 2 | 0.2 | 4 | 0.5 | 0.22 |

1) Distance measured from the impacted surface of the primary target (see Fig. 1).
2) Angle between the primary target surface and the direction to the center of the secondary target (see Fig. 1).
3) Framing interval.
4) Exposure duration.
5) Length of a vertical pixel side on the glass plate.

The primary targets were basalt blocks with each side at least 100 mm from Kinosaki, Hyogo Prefecture, Japan. The average density was 2700 kg $m^{-3}$. The previously measured compressional- and shear-wave velocities and the uniaxial compressive strength of the material were 4.64 km $s^{-1}$, 2.93 km $s^{-1}$, and ~220 MPa, respectively (Matsui et al. 1982). We used the Brazilian disk test (Nakamura et al. 2007) to measure the static tensile strength of basalt disks 8.9 mm in diameter and 6.1 mm thick, respectively; the result was 12.0 ± 5.6 MPa. Three adjacent tempered glass plates were placed as secondary targets at a distance of approximately 400 mm ($r_0$ in Fig. 1a) from the impacted point of the primary target. Each glass plate had a density of 2500 kg $m^{-3}$, with dimensions of 100 × 50 × 5 mm (height × width × thickness). We chose tempered glass instead of ordinary glass because it is easier to handle; when broken, it fragments into fine granules rather than sharp pieces. Before the shot, grid lines with approximately 10 mm spacing were drawn on the glass plates using an oil-based marker. The glass plates were fixed to a metal frame with tape along their top and bottom edges, as shown in Fig. 1b. Since the tape covered approximately 5 mm along both the top and bottom edges in the vertical direction, the effective size of the secondary target (i.e., the area exposed to the ejecta) was approximately 90 mm in height and 150 mm in width.

The angle between the primary target surface and the normal at the center of the secondary target ($\theta_0$ in Fig.1a) was approximately 60°. The ejection angle is defined as the angle between the primary target surface and the ejection direction ($\theta$ in Fig.1a). The detection range covered by the secondary target was approximately 45–70°. The incident angle of the ejecta on the glass plates, $\beta$ in Fig.1b, is given by

$$\beta = \tan^{-1}\left(\frac{r_0}{l}\right),$$

where $l$ is the distance between the impact point of the ejecta and the center of the secondary target. Considering that $l <$ 87.5 mm (reduced due to the taped area of the glass), the possible range of $\beta$ were 77°–90°, 79°–90°, and 78°–90° for shots #4982, #5115, and #5231, respectively.

Optical microscope images of the secondary target, taken at different resolutions (details are provided in Section 3), were obtained after each experiment. The entire system was mounted in a vacuum chamber with acrylic resin windows, the surfaces of which were either parallel to or at 45° to the projectile trajectory. The ambient chamber pressure was ~2 Pa.

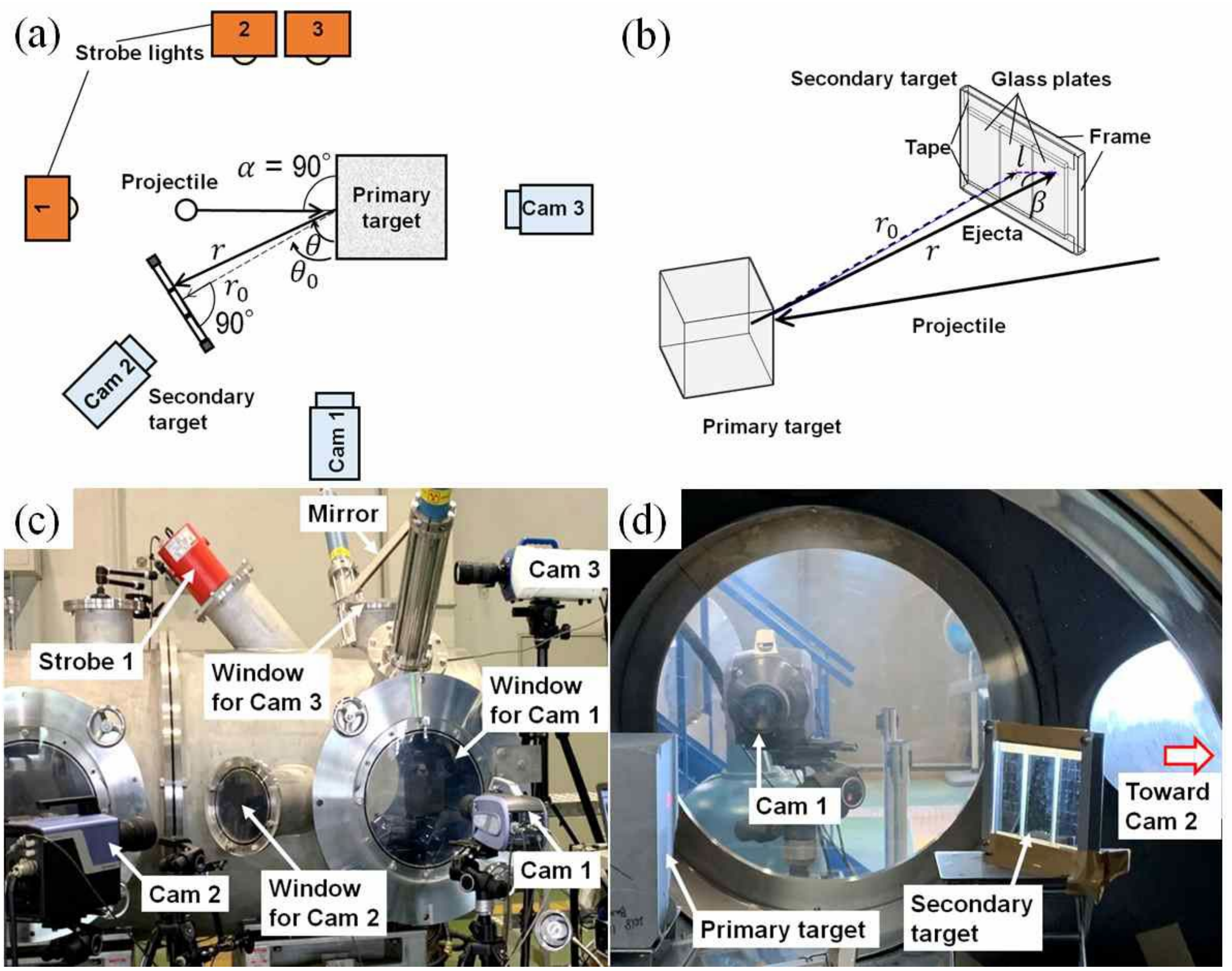

Fig. 1. Experimental setup. (a) Schematic view from the direction vertical to the projectile trajectory. Strobe 1 was mounted on an obliquely downward-facing window above the chamber and illuminated the target at a 45° angle from above. (b) Schematic view of the ejecta trajectory and the secondary target. (c) Outside view of the chamber from the side of the cameras (Cam 1 and Cam 2). The mirror (omitted from Fig. 1a) was mounted at a 45° angle above the upper window of the chamber and was used by Cam 3 to capture images of the primary target from above. (d) Inside view of the chamber from the direction opposite to that shown in (c).

The projectile impact on the primary target was recorded by a high-speed camera (HPV-2 or HPV-X) on the side of the primary target (below, "Cam 1") at 2-μs intervals. The impacts of ejecta on the secondary target were captured by one or two high-speed cameras (HPV-X, HPV-2, or MEMRECAM ACS-1) ("Cam 2" below), from an oblique upstream direction through a window mounted on the chamber at a 45° angle. The frame rates were set at 30,000, 100,000, or 250,000 frames $s^{-1}$, i.e., at 33-, 10- or 4-μs intervals ($T_{\mathrm{int}}$), with exposure durations ($T_{exp}$) of 2, 8, or 0.5 μs. Another high-speed camera (Kirana), referred to as "Cam 3," was mounted horizontally and captured downward-looking images through a mirror tilted at 45° and a window at the top of the chamber. It also recorded impact positions at 2- or 5-μs intervals to check the projectile trajectory in the horizontal plane, which cannot be determined from the Cam 1 images. We used three strobe lights to provide the illumination for the cameras. One strobe illuminated the chamber through a diagonally mounted window at the top of the chamber, and two shone through a window opposite the position of Cam 1; Cam 1 thus operated under backlit lighting. All cameras and strobes were controlled by a single trigger emitted by the laser velocimetry system. However, there was no control or monitoring of when, in each framing interval, the camera would initiate the exposure.
As shown in Table 1, the spatial resolution of Cam 2 was better than 0.5 mm pixel$^{-1}$. With this setup, the impact points of the high-velocity ejecta on the secondary target, typically spaced about 1 mm to a few millimeters apart, could be captured in the high-speed camera images. Because a single pixel in the high-speed images rarely contained more than one candidate impact point (except for doublets, referred to later in the text), it was possible to match the impact points observed in the high-speed camera images with the craters on the secondary target identified in the microscope images taken after the experiment. The procedure for matching the impact points on the glass plates in the high-speed camera images with the corresponding craters in the microscope images is explained in Section 3.1.

### *2.2. Crater Scaling Relations for the Secondary Target*

We conducted a series of calibration experiments to establish the scaling relationship between the diameter of the crater formed on the secondary target (the glass plate) and the size of the ejecta particle that formed the crater. In these experiments, the projectiles were soda-lime glass spheres (diameter, 0.324 ± 0.01 mm; density, 2520 kg m$^{-3}$). The impact velocity ranged from 2 to 7 km s$^{-1}$ (Table A1). Images of craters on the glass plates were captured using an optical microscope.
Figures 2a and 2b show craters on the glass plates from the calibration shots. The outermost side of each crater exhibits a thin fissure with arcuate stripes. At velocities of ~3 to 7 km s$^{-1}$, each crater had a distinct, circular central zone (probably a pit) surrounded by an uneven outer region, as in a previous study using soda lime glass blanks (Burchell and Gray, 2001).

However, the central circular zone was absent in craters formed at a velocity of ~2 km s$^{-1}$. We used ImageJ software (Image Processing and Analysis in Java: https://imagej.nih.gov/ij/) to measure the equivalent circle diameters of the boundary between the uneven outer region and the outermost fissure region. As the precise boundaries were difficult to discern, we measured the minimum and maximum diameters and averaged them.

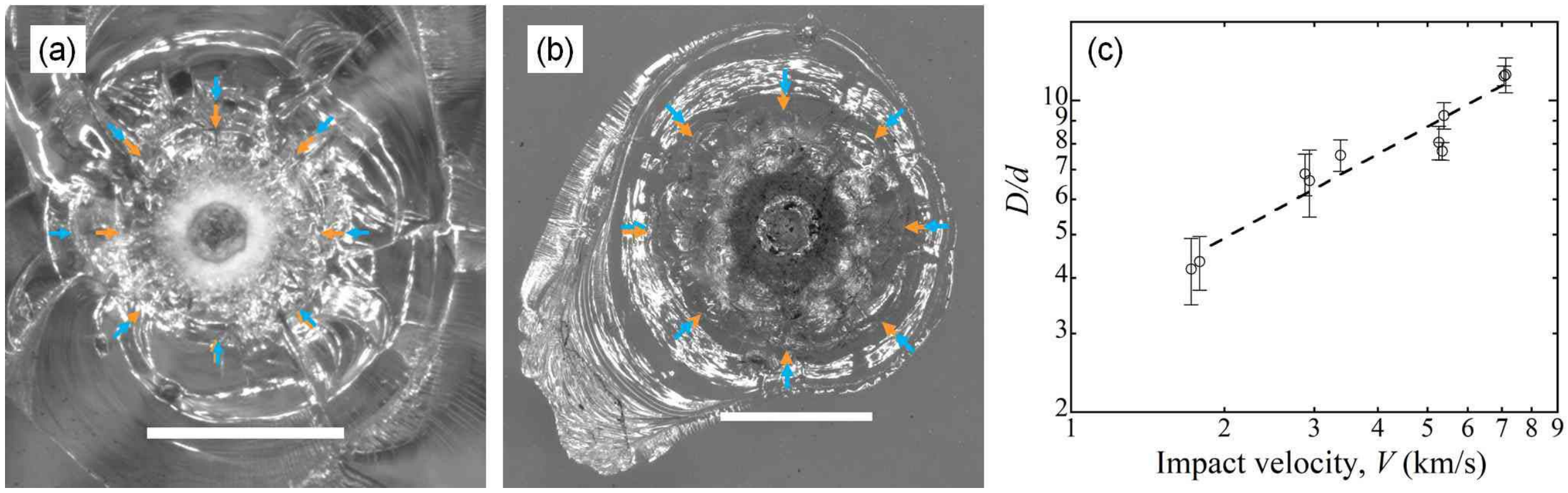

Fig. 2. Optical microscope images of craters formed during calibration shots at impact velocities of (a) 3.38 km s$^{-1}$ and (b) 7.07 km s$^{-1}$. Blue and orange arrows indicate the outermost and innermost boundaries between the uneven outer region and the outermost fissure region, corresponding to the maximum and minimum diameters, respectively (see main text for details). Scale bars: 2 mm. (c) Normalized crater diameters (circles) at each impact velocity. The two ends of the error bar correspond to the maximum and minimum values. The dashed line shows the empirical relationship of Eq. (2).

To estimate the ejecta particle size using the crater size on the secondary target (the glass plate), we analyzed the results using the crater diameter π-group scaling relationship of the strength regime (e.g., Holsapple and Schmidt, 1982). This relationship uses the dimensionless parameters $\pi_{\mathrm{D}} = (\frac{\rho}{m})^{\frac{1}{3}}D$, $\pi_3 = \frac{Y}{\delta V^2}$, and $\pi_4 = \frac{\rho}{\delta}$ and the coefficients $C_1$, $C_2$, and $C_3$:

$$\pi_D = C_1 {\pi_3}^{C_2} {\pi_4}^{C_3} \qquad (1)$$

where $\delta$, $m$, and $V$ are the density, mass, and velocity of the impacting particle, respectively; and $\rho$, $Y$, and $D$ the target density, strength, and crater diameter, respectively. The ratio of the crater diameter $D$ to the equivalent spherical diameter of the impacting particle $d$ is:

$$\frac{D}{d} = C_4 V^{-2C_2} \qquad (2)$$

where:

$$C_4 = C_1 \left(\frac{\pi}{6}\right)^{\frac{1}{3}} {\rho_{\mathrm{gp}}}^{\left(C_3 - \frac{1}{3}\right)} {Y_{\mathrm{gp}}}^{C_2} {\delta_{\mathrm{gs}}}^{\left(\frac{1}{3} - C_2 - C_3\right)} \quad (3)$$

Here $\rho_{\mathrm{gp}}$, $Y_{\mathrm{gp}}$, and $\delta_{\mathrm{gs}}$ represent the density of the glass plate (target), strength of the glass plate, and the density of the glass sphere (projectile). Figure 2c shows the diameter ratio versus the impact velocity (km s$^{-1}$). Although the data are scattered, the average crater size is proportional to a power of the impact velocity. The fitted parameters in Eq. (2) were $C_2$ = −0.31 ± 0.03 and $C_4$ = 3.2 ± 0.3, respectively. The relationship between the equivalent sphere diameter of basalt ejecta particle $d_{\mathrm{basalt}}$ that forms a crater of equivalent circle diameter $D_s$ on the glass plate is given by:

$$\frac{D_{\mathrm{s}}}{d_{\mathrm{basalt}}} = C_4' {V_{\mathrm{ej}}}^{-2C_2} \quad (4)$$

where:

$$C_4' = C_1 \left(\frac{\pi}{6}\right)^{\frac{1}{3}} {\rho_{\mathrm{gp}}}^{\left(C_3 - \frac{1}{3}\right)} {Y_{\mathrm{gp}}}^{C_2} {\delta_{\mathrm{basalt}}}^{\left(\frac{1}{3} - C_2 - C_3\right)} \quad (5)$$

Here $\delta_{\mathrm{basalt}}$ and $V_{\mathrm{ej}}$ represent the density of the basalt ejecta, and the ejecta velocity impacting the glass plate, respectively. By combining Eqs. (3)–(5), we obtain a relationship between the crater diameter on the glass plate ($D_s$) and the basalt ejecta particle diameter ($d_{\mathrm{basalt}}$), in which the target strength is not involved:

$$\frac{D_{\mathrm{s}}}{d_{\mathrm{basalt}}} = C_4 \left(\frac{\delta_{\mathrm{basalt}}}{\delta_{\mathrm{gs}}}\right)^{\left(\frac{1}{3} - C_2 - C_3\right)} {V_{\mathrm{ej}}}^{-2C_2} \quad (6)$$

We assumed $C_3$ = 0.11, as in previous work using sedimentary rock targets (Suzuki et al. 2012); this small value implies that ${\pi_4}^{C_3}$ in Eq. (1) is approximately equal to unity. As explained in Section 2.1, the incident angle of the ejecta on the secondary target is nearly perpendicular ($\beta$=77°–90°). Impact cratering experiments on glass using particles incident at oblique angles showed that the crater length (the dimension along the line of flight) and width (the dimension orthogonal to the length) were not affected by incident angle for $\beta$=45°–90° and $\beta$=20°–90°, respectively (Burchell & Grey, 2001). Therefore, we did not apply any correction to the incident angle.

## 3. Analyses

### *3.1. Ejecta Detection using the Secondary Target*

On the basalt target, a crater with a spall region approximately 40 mm in diameter and a central pit was formed, as shown in Fig. 3a. The size of craters formed under similar conditions is discussed in detail elsewhere (Kadono et al., 2023). Figure 3b is a side-view image from Cam 1 obtained immediately after a projectile struck the basalt target. Figures 3c and d are Cam 1 images taken 2 and 66 μs later. Figures 3e-g are Cam 2 images of the glass plates taken from a direction oblique to the projectile trajectory. The projectile approaches the basalt target from left to right. Two of the three plates (middle and right) are shown here in the cropped image in Figures 3e-g. The third plate (on the left) is not shown here so that the two displayed plates can be enlarged, although it does appear in the original camera images.

In the right of Fig. 3f, part of the ejecta cone is shown. The projectile impacted the basalt target after the image shown in Fig. 3e had been acquired. In Fig. 3f, the ejecta cone is more developed (extended) than in Figs. 3b and c. Thus, the projectile impacted the basalt target well before the end of the 2-μs exposure of Fig. 3f. In Figs. 3g-i, the bright spots in the images show the impacts of high-speed ejecta near the boundary of the two glass plates. In general, the impacts caused either a sudden increase or decrease in the pixel grayscale. This change in grayscale is likely due to the way light is scattered by cracks formed within the glass as a result of crater formation. The cracked areas appear either brighter or darker than the intact surrounding regions and retain this appearance over time, until subsequent impacts create new cracks that modify the light scattering state. Some pixels appear brighter in only one frame, which may be attributed to scattered light from numerous micro-ejecta captured at the precise moment of their ejection from the secondary target. Even when an ejecta particle impacts the secondary target, it cannot be detected unless it produces a discernible change in the grayscale of the high-speed camera images. Together with the limited solid angle covered by the secondary target, this means that the ejecta analyzed in this study represent only a small fraction of the total.

Figure 3j illustrates the Cam 2 operation sequence and the events captured for shot #4982. The projectile impacted the basalt target after the image in Fig. 3e had been taken but before the end of the exposure of the image in Fig. 3f, i.e., at $T_1$. The ejecta impacted the glass plates (Fig. 3g) at $T_2$.

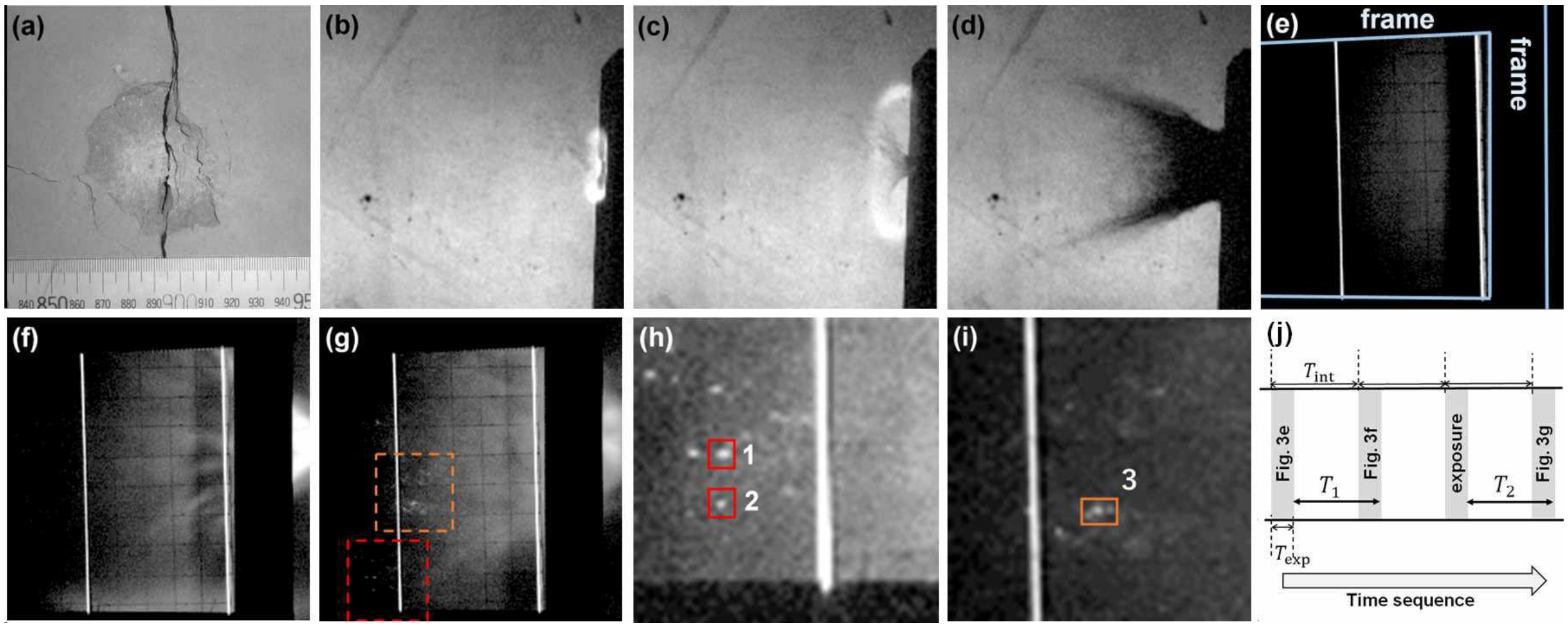

Fig. 3. Results for one shot (#4982). (a) The crater on the basalt target. (b)–(d) Images taken by Cam 1, and (e)–(i) by Cam 2. All images are partially cropped, brightness-adjusted versions of the original images. In (e)–(g), the two vertical white lines represent the edges of the glass plates, and the black grid on the glass plates have approximately 10 mm spacing. The light blue lines in (e) delineate the metal frame supporting the glass plates (see Fig. 1d). (b) Cam 1 image obtained just after the projectile impacted the basalt target. (c) The subsequent Cam 1 image, taken 2 μs later. (d) The ejecta cone captured by Cam 1 taken 66 μs after (b). (e) Cam 2 image obtained just before the projectile struck the basalt target. (f) Image taken one frame after (e), i.e., 33.3 μs later. (g) Image taken two frames after (f), i.e., 66.7 μs later. (h) Enlarged and enhanced view of the square outlined by the red dashed lines in (g). The bright spots (numbered 1 and 2) are shown in Figs. 4a–c. (i) Enlarged and enhanced view of the rectangle outlined by the orange dashed lines in (g). The bright spots (numbered 3) are shown in Fig. 4d. (j) Time sequence of the Cam 2 operation and the event. The projectile impact occurred at $T_1$, and the fastest ejecta impacted the secondary target at $T_2$.

Bright/dark spots in the Cam 2 images were first manually matched to craters in the low-resolution optical microscope images (11 μm pixel$^{-1}$ or better). When ambiguous, only those matches in which the relative coordinates (determined essentially by affine transformation) of the craters were consistent within approximately one pixel in the high-speed camera images were retained.

Figure 4 shows optical microscope and scanning electron microscopy (SEM) images of the secondary target. Notably, ejecta impact points with craters having a 'diameter' (based on our definition) smaller than one pixel of Cam 2 images were identified. It is possible that the cracks in the glass extend beyond the size of a single pixel in the high-speed camera, which may have contributed to the observed change in brightness. In the high-resolution microscope images (0.66 μm $pixel^{-1}$ or better; Figs. 4b–4d, the insets of Fig. 4d, and 4f), the craters consist of an uneven region surrounded by thin fractures. The morphology of the craters on the secondary targets was similar to those formed in the calibration shots described in Section 2.2, in which the projectile was a solid glass particle. Owing to this morphological similarity between the calibration shots and the secondary targets, the ejecta size can be estimated from Eq. (6) once the ejecta velocity ($V_{ej}$) is known (see Section 3.2). The equivalent circle diameters of the boundaries between the uneven region and the outermost fissure were measured for those that could be identified with the bright/dark spots in the Cam 2 images. Some craters observed as doublets (Fig. 4b), potentially formed by fragments that broke during flight or by ejecta launched at different times on similar trajectories. Only the diameter of the larger crater was measured.

We subjected the left glass plate in shot #4982 to SEM/energy-dispersive x-ray spectrometry (EDS); we sought, but did not find, ejecta material in craters. However, we observed several aluminum-rich splash patterns attributable to the aluminum impacts, with examples shown in Fig. 4e. Therefore, the craters on the glass plates in the shots using aluminum projectile (#4982 and #5231) were likely formed by the basalt (target) ejecta, rather than the ejecta from the aluminum projectiles.

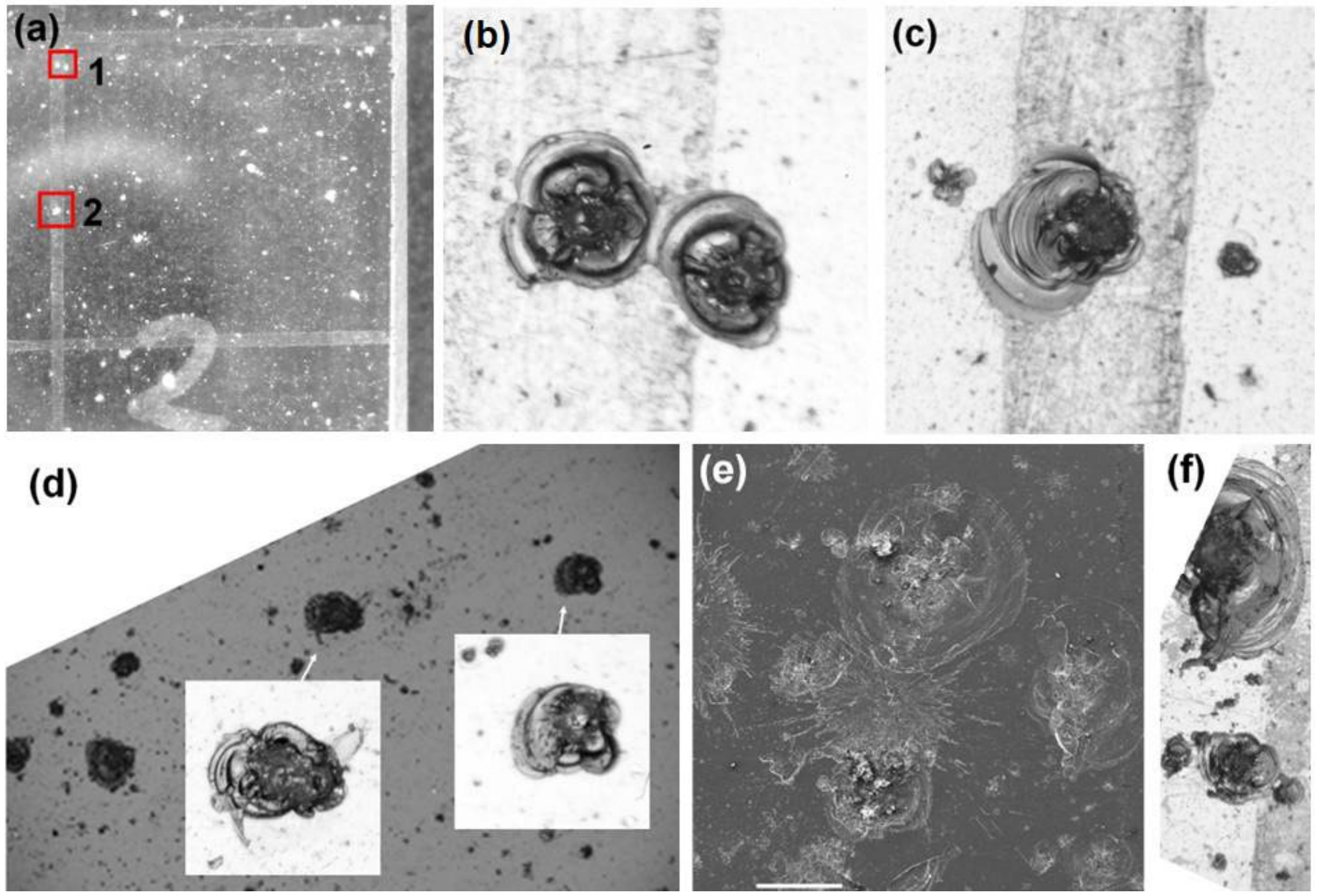

Fig. 4. Image of the glass plates (shot #4982). (a) A low-resolution optical microscope image of craters in the central plate. The bright spots numbered 1 and 2 are those in Fig. 3h. The grid with approximately 10 mm spacing that appears black in Figs. 3e–g appears white in Fig. 4a. The white letter “2” on the plate indicates that this is the central plate in the three-plate set. (b) Enlarged view of doublet craters 1 in (a). Each side of the image is 0.7 mm in length. (c) Enlarged view of crater 2 in (a). Each side of the image is 1 mm in length. (d) Optical microscope image (2.8 mm wide) showing craters on the right glass plate, corresponding to Area 3 in Fig. 3i. The insets show magnified views, each 0.35 mm on a side. (e) SEM image (secondary electron image) of radially patterned aluminum deposits on the left glass plate. Scale is 0.5 mm in length. (f) An optical microscope image corresponding to the bottom-right region of the SEM image in (e).

### *3.2. Estimation of the Ejection Velocity*

Using the images of Cam 1 and Cam 2, the time between the primary and secondary impacts was estimated:

$$T_{\mathrm{lag}} = T_2 - T_1 \qquad (7)$$

Here, $T_1$ is the time at which the projectile impacted the basalt target, and $T_2$ is the impact time when the ejecta impacted the secondary target. For example, for the shot in Figs. 3 (#4982), $T_{\mathrm{lag}}$ for the fastest ejecta detected in Fig. 3g is $T_{\mathrm{int}} < T_{\mathrm{lag}} < 3T_{\mathrm{int}}$. Other shots (#5115 and #5231) were similarly analyzed. A numerical approach revealed that the time interval between the ejecta launch and the moment of impact, for ejecta with velocities exceeding the impact velocity, was less than $d_{\mathrm{p}}/(V_{\mathrm{p}}\sin\alpha)$ (Wakita et al. 2021), where $d_{\mathrm{p}}$ and $V_{\mathrm{p}}$ are the projectile diameter and impact velocity, respectively. Here, $d_{\mathrm{p}}/(V_{\mathrm{p}}\sin\alpha)$ ~0.5 μs is much smaller than the travel time of the ejecta with a velocity of ~7 km $s^{-1}$ from the primary target to the secondary target, i.e., ~60 μs. We thus neglected any time interval between the projectile impact on the basalt target and ejecta launch.

The ejecta velocity was estimated using the following equation:

$$V_{\mathrm{ej}} = \frac{r}{T_{\mathrm{lag}}} \qquad (8)$$

where $r$ is the distance between the impact point on the primary target and the center of the crater formed on the secondary target (Figs. 1a and 1d). In this study, the flight distance of the ejecta was approximated by $r$. To evaluate the uncertainty associated with this approximation, we referred to the numerical simulation results presented in Kurosawa et al. (2018), specifically the 6 km $s^{-1}$ and 12 km $s^{-1}$ cases shown in Fig. S10, among others. According to these simulations, target materials that experienced pressures comparable to those estimated for Martian meteorites (30–50 GPa) and were ejected at velocities exceeding 5 km $s^{-1}$ originated from very specific regions: radial distances within about one projectile radius from the impact axis and from extremely shallow layers less than 2% of the projectile radius below the target surface. In our experiments, the projectile radius (1.6 mm) was sufficiently small compared with the distance between the impact point and the secondary target (≈ 400 mm). Therefore, the error associated with the ejection position of high-velocity ejecta (>5 km $s^{-1}$) that experienced pressures of 30–50 GPa is considered negligibly small. The ejecta analyzed in this study, which were launched at velocities ranging from 1 to 7 km $s^{-1}$ and experienced a wider range of pressures, are inferred to have been ejected from a broader radial range than the aforementioned high-velocity ejecta. However, these ejection points likely did not extend to the rim of the crater formed in the basalt primary target (radius ≈ 20 mm, i.e., 1/20 of $r$≈400 mm), because ejecta originating from the outermost rim are slow fragments produced during the final stage of crater formation (e.g., Housen and Holsapple, 2011). For these reasons, the error associated with approximating the ejecta flight distance by $r$ is estimated to be less than a few percent, and thus much smaller than the uncertainty in $T_{\mathrm{lag}}$. We estimated the maximum possible $V_{\mathrm{ej,max}}$ and minimum possible $V_{\mathrm{ej,min}}$ velocities using the minimum and maximum values for $T_{\mathrm{lag}}$.

## 4. Ejecta Velocities and Comparison with Previous Results

Using Eq. (8) to derive the ejecta velocities, and applying Eq. (6), we estimated the sizes of the largest ejecta in several Cam 2 frames for the three shots in Table 1. Only the largest ejecta impacting the secondary targets can be detected in the high-speed images. For example, the two smaller scars on the left and right of the central crater measured in Fig. 4c probably formed at a similar time as the central crater; one larger and two smaller ejecta shared a near-identical ejection velocity. The largest ejecta at each ejection velocity allow simultaneous measurement of ejecta size and velocity. We also estimated the ejection angle from the basalt target surface ($\theta$).

Figure 5 shows the size–velocity and ejection angle–velocity relationships for the largest ejecta. Because the framing interval is shorter in shots #5115 and #5231 than in shot #4982, and the distance between the primary and secondary targets is greater, the velocity estimations are more reliable in the second and third shots; the ejection velocities are less ambiguous. Figure 5a shows the average of $V_{\mathrm{ej,max}}$ and $V_{\mathrm{ej,min}}$ values; the differences $(V_{\mathrm{ej,max}} - V_{\mathrm{ej,min}})/2$ are indicated by the error bars. However, in Figs. 5b-d, the plot of $V_{\mathrm{ej,min}}$ versus the corresponding ejecta sizes is plotted as a conservative estimate of the ejection velocity. No significant differences between projectile material were detected in the upper envelope curve of the size–velocity relationship or in the general trend of the ejection angle-velocity relationship. In this experiment, the maximum ejecta velocity captured was higher for the alumina projectile. This may be because the initial pressure was greater for the alumina projectile (with high impedance) than for the aluminum one.

Figure 5c shows that the ejection angle increases with a decrease in velocity at ejection velocities above ~2 km $s^{-1}$, which contrasts with typical ejecta behavior in normal excavation flow, where higher ejection angles are associated with higher velocities, as seen when aluminum projectiles impacted basalt targets at velocities of 6.1 to 6.4 km $s^{-1}$ (Gault et al. 1963). The blank areas (no data points) in the bottom left of Figs. 5a, b, and d are attributable to measurement bias. We suggest the existence of ejecta of the size and ejection velocity that would fill the blank area in the figures.

We detected ejecta tens of microns in diameter with ejection velocities up to several km $s^{-1}$, exceeding the escape velocities of the Moon, Mercury, and Mars.

Martian meteorites acquired ejection velocities exceeding the escape velocity of Mars (>5 km $s^{-1}$), while the maximum pressures they experienced did not reach levels sufficient to completely melt the rock. A mechanism proposed to enable such ejection is spallation—the complex interaction between shock waves and rarefaction waves reflected from the surface—which expels solid material at high velocity (Melosh, 1984; Head et al., 2002; Artemieva and Ivanov, 2004; Kurosawa et al., 2018; Bowling et al., 2020; Elliott et al., 2022). An analytical model describing this mechanism indicates the following relationship between ejection velocity and fragment thickness for spall fragments $z_\mathrm{s}$ with an ejection velocity $v_\mathrm{e}$ ≤ approximately 1 km $s^{-1}$ (Melosh 1984):

$$z_s \sim \left(\frac{\sigma_\mathrm{t}}{\rho c_\mathrm{L}}\right)\frac{d}{v_\mathrm{e}} \qquad (9)$$

where $\sigma_\mathrm{t}$ and $c_\mathrm{L}$ are the target tensile strength and longitudinal wave velocity, respectively. Figure 5b extrapolates the model slightly beyond $v_\mathrm{e}$ ≈ 1 km $s^{-1}$ for reference, although the point-source approximation used to derive Eq.(9) is no longer valid in this regime. As for the tensile strength, because laboratory studies have shown that the dynamic tensile strengths of rocks are several-to ten-fold greater than the static values (e.g., Grady & Kipp 1980; Padmanabha et al. 2023), the model curves in Fig. 5b correspond to tensile strengths that are 10- and 100-fold greater than the static tensile strength of the basalt used in this study. On the other hand, numerical simulations examining the spallation process—namely, the release of ejecta resulting from the complex interaction between shock and rarefaction waves—have shown that ejecta with velocities exceeding 5 km $s^{-1}$ and pressures below 50 GPa (lower than the pressure required for complete rock melting; Stöffler et al., 2018) can be launched from near-surface regions of the target (e.g., Kurosawa et al., 2018; Bowling et al., 2020; Elliott et al., 2022). The high-velocity ejecta detected from the craters on the secondary target in this study likely correspond to such solid ejecta.

The impact velocity in this study, ~7 km $s^{-1}$, is lower than the average impact velocity of meteoroids on the Moon, which exceeds 10 km $s^{-1}$ (e.g., Bottke et al. 1994). However, it has been shown that some meteoroids larger than 10 cm impact the Moon at velocities lower than 10 km $s^{-1}$ (Marchi et al. 2009). On Mercury, the impact velocity of large meteoroids, which are not influenced by the Sun's radiative forces, is even higher, ranging from about 15 to 80 km s$^{-1}$ (Marchi et al. 2005; 2009). In contrast, smaller meteoroids, which are significantly affected by the Sun's radiative forces, tend to have different impact velocities (e.g. Borin et al. 2017): for example, the impact velocity of a meteoroid with a diameter of 10 μm on Mercury is estimated to range between 5 and 30 km s$^{-1}$ (Pokorný et al. 2018). The results of this study suggest that fragments can be ejected into interplanetary space as a result of meteoroid impacts on the rocky surfaces of the Moon and Mercury. Further experimental studies are needed to investigate the dependence of impact velocity, impactor size, and the materials of both target and projectile on the ejecta size–velocity relationship.

Figure 4d shows the experimental size–velocity relationship of ejecta from basalt targets over a wide velocity range. The lower-velocity results (from a few m s$^{-1}$ to several tens m s$^{-1}$) were obtained when projectiles obliquely impacted spherical, rather than flat, targets. It should be noted that both the results of this study and those of previous research represent the velocities of the largest ejecta measured within their respective ejection velocity ranges. Although ejecta smaller than those shown in the figure and having comparable velocities are likely to exist, they have not been measured. It should also be noted that the lack of overlap between the results of previous studies and the present findings stems from differences in measurement methods, specifically, the ejection velocity ranges targeted for measurement. Although the data for ejecta with velocities between 100 m s$^{-1}$ and 1 km s$^{-1}$ are lacking, the plot in Fig. 5d suggests that the maximum ejection velocity decreases monotonically with increasing ejecta size.

## 5. Summary

We used high-speed camera and microscope images to simultaneously estimate the velocities and sizes of the largest high-velocity ejecta generated by vertical impacts of aluminum and alumina projectiles onto basalt targets at 7 km s$^{-1}$. We defined the size-velocity relationships for sub-millimeter ejecta with velocities > 1 km s$^{-1}$. The quantitative results of this study and subsequent investigations into the dependence of the ejecta size–velocity relationship on impactor size, impact velocity, and material using the novel method presented here, can contribute to future studies on the supply of planetary material from solid surfaces, such as those of Mercury and the Moon into interplanetary space. They can also be compared with the results of models and numerical simulations (e.g., Kurosawa et al., 2018; Bowling et al., 2020; Elliott et al., 2022) that seek to understand high-velocity impact ejecta, including those responsible for the ejection of Martian meteorites.

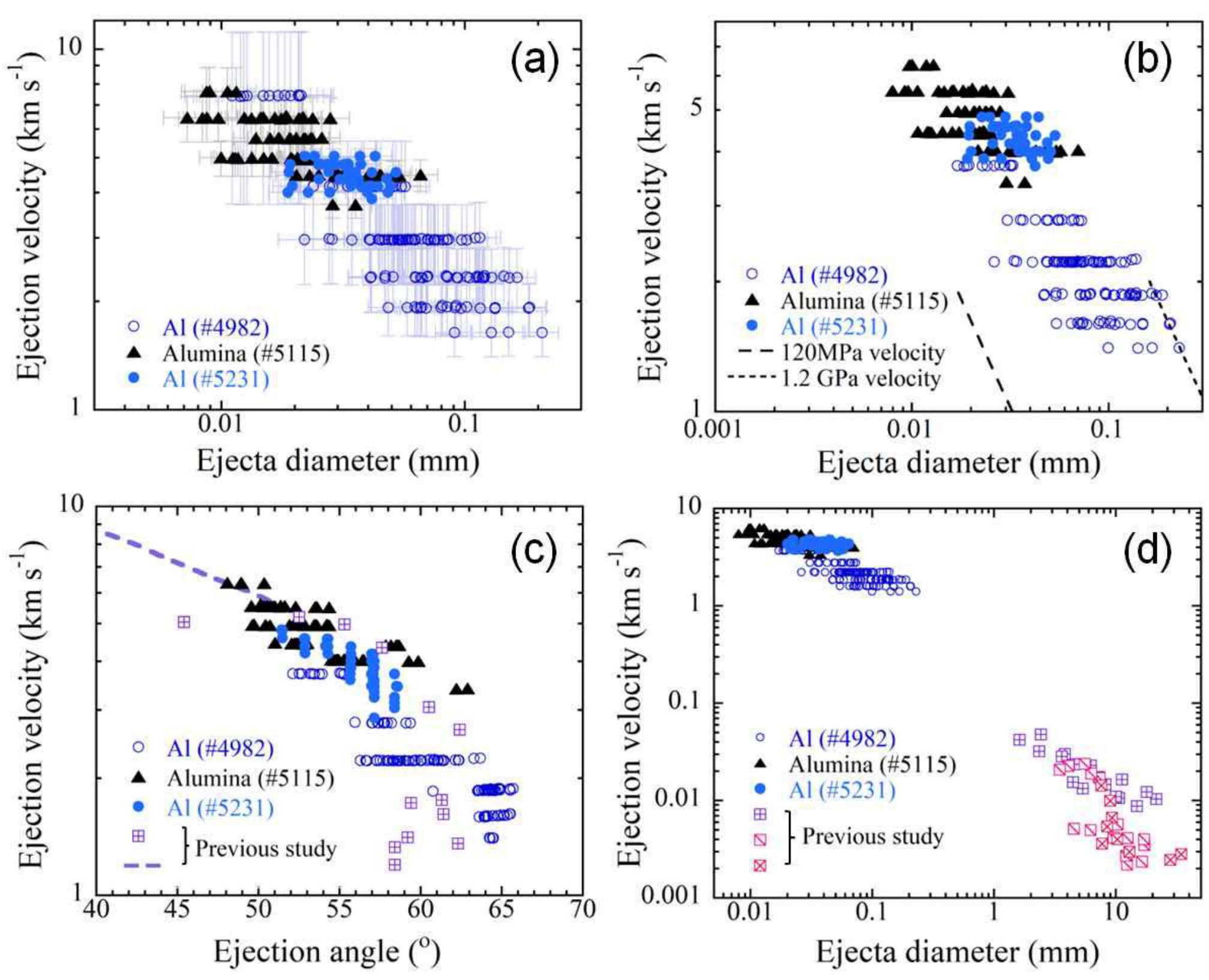

Fig. 5. Relationship between ejecta size, velocity, and ejection angle. (a) Relationships between ejecta size and ejection velocity: $(V_{\mathrm{ej,max}} + V_{\mathrm{ej,min}})/2$. The error bars correspond to $(V_{\mathrm{ej,max}} - V_{\mathrm{ej,min}})/2$. The ejecta size is represented on the horizontal axis in terms of the equivalent sphere diameter. The error bars include a 10% measurement uncertainty in crater size, together with the estimated errors in velocity and the uncertainty of the scaling law. (b) $V_{\mathrm{ej,min}}$ versus ejecta size derived from Eq. (6), where $V_{\mathrm{ej}} = V_{\mathrm{ej,min}}$. The dashed and dotted curves are the extrapolations of an analytical model [Eq. (9)] assuming that the tensile strengths are 10- and 100-fold that of the static tensile strength, respectively. (c) Relationship between ejection angle and velocity ($V_{\mathrm{ej,min}}$). The dashed curve and square marks are from a previous study (Gault et al. 1963). (d) Relationship between ejecta size and velocity ($V_{\mathrm{ej,min}}$). Previous results of impacts with oblique incidences of 30°, 45°, and 60° of basalt targets are shown by squares containing a plus sign, diagonal line, and a cross, respectively (Nakamura & Fujiwara 1991; Nakamura 1993).

Acknowledgments

We thank Y. Seto (Osaka Metropolitan University) and A. Iemoto (Kobe University) for the technical assistance regarding SEM-EDS measurements. This research was supported by JSPS KAKENHI Grant Numbers JP20K04055 and JP24K00695 and the Hypervelocity Impact Facility at ISAS/JAXA. In revising the manuscript, the authors used ChatGPT-5 solely for the purpose of English proofreading. After using the tool, they carefully reviewed the content and made revisions as necessary.

## Appendix

Table A1 summarizes the data obtained for each calibration shot.

Table A1. Impact velocities and crater diameters for the calibration shots.

| Shot# | Impact velocity, $V$ (km s$^{-1}$) | Normalized Crater diameter, $D/d$ | |
|---|---|---|---|
| | | Min. | Max. |
| Cal-1 | 3.38 | 6.93 | 8.17 |
| Cal-2 | 5.26 | 7.37 | 8.75 |
| Cal-3 | 5.35 | 7.35 | 8.05 |
| Cal-4 | 5.38 | 8.64 | 9.89 |
| Cal-5 | 2.87 | 6.12 | 7.58 |
| Cal-6 | 2.93 | 5.48 | 7.74 |
| Cal-7 | 1.79 | 3.75 | 4.96 |
| Cal-8 | 7.12 | 10.4 | 12.5 |
| Cal-9 | 1.72 | 3.48 | 4.90 |
| Cal-10 | 7.07 | 10.8 | 12.0 |